\documentclass[]{aa}

\usepackage{graphics} 
\usepackage{amssymb}
\begin{document}

\author{Sune Toft \inst{1} 
\and Jens Hjorth  \inst{1}
\and Ingunn Burud \inst{2}}

\institute{Astronomical Observatory, University of Copenhagen, Juliane
  Maries Vej 30, DK--2100 Copenhagen \O, Denmark 
\and Institut d'Astrophysique et de G\' eophysique, Universit\' e de
  Li\`ege, Avenue de Cointe 5, B--4000 Li\`ege, Belgium} 

\title{The extinction curve of the lensing galaxy of
  \object{B1152+199} at $z=0.44$\thanks{Nordic Optical Telescope is operated on the island of La Palma jointly
by Denmark, Finland, Iceland, Norway, and Sweden, in the Spanish Observatorio del Roque de los Muchachos of the Instituto de Astrofisica de Canarias.}}

\thesaurus
{03(11.17.4 \object{B1152+199}; 11.09.4; 09.04.1; 09.05.1; 12.07.1) }

\titlerunning{The extinction curve of the lensing galaxy of \object{B1152+199}}

\maketitle  
\abstract We present  $UBVRIz^\prime$ photometry  of 
the gra\-vitational lens candidate CLASS \object{B1152+119} obtained  with  the  Nordic  
Optical Telescope.  The two QSO components  are resolved in the $B$, $V$, $R$, $I$ 
and $z^\prime$ bands confirming the lensing nature of the system. 
The $z=0.44$ lens galaxy is clearly detected in $B$,  
$R$, $I$ and $z^\prime$ and its position is found to be almost  coincident 
with the faint  QSO image which is heavily extincted (relative to 
the brighter  QSO image) by dust in  the lens  galaxy.  The extinction  
curve of the   lens galaxy derived from the relative   photometry  is 
well fitted by a Galactic extinction  law with  $1.3 \lesssim R_V
\lesssim 2.0$ and
$E(B-V) \approx 1$. From a simple model of the system we predict a
time delay of $\sim$ 60 days.

\keywords{quasars: individual: \object{B1152+199} -- galaxies: ISM -- ISM: dust, extinction -- ISM:
  evolution -- gravitational lensing }

\section{Introduction}
Dust  is ubiquitous in the  Universe and is responsible for extinction
of the light  from  distant sources.   The type of  dust (composition,
grain  sizes and shape)  determines  the  amount  of extinction as   a
function of wavelength  --  the extinction  curve $A_{\lambda}$.  
Extinction  curves
almost certainly evolve with redshift since the metallicity, elemental
abundance ratios, mean star-formation rate, and energy-injec\-tion rates
that determine the structure and evolution of the  dust are all strong
functions  of redshift.  The  extinction in our   Galaxy has been well
studied   (Cardelli et  al. 1989),  but  very  little  is known  about
extinction in  other galaxies, especially at  higher redshifts. This is
unfortunate as   dust  plays  an   increasingly important   role    in
contemporary cosmology.

%A variety of techniques has been attempted to determine extinction
%curves of external galaxies. Goudfrooij et al. (\cite{goudfrooij})
%showed that the extinction laws of large dust rings in nearby
%elliptical galaxies also follow the Galactic extinction law with
%$R_V$ ranging from 2.1 to 3.3. This method is limited to low-$z$
%galaxies with well-defined elliptical isophots, and hence can not be
%applied to galaxies of irregular or late-type morphology or galaxies
%at moderate or high redshift.

%Gravitationally lensed images are expected to have identical colors
%due to the achromatic nature of gravitational lensing. Any color
%difference between the QSO images can therefore be interpreted as differential extinction of the images.
Recently the use of multiply imaged QSOs has been explored as a means
of inferring differential extinction curves for distant galaxies
(Nadeau et al.~\cite{nadeau}; Malhotra et al.~\cite{malhotra}; 
Falco et al.~\cite{Falco}). The
idea behind this approach is that a lensed QSO can be used as a
`standard beacon', shining through different paths of
the lensing galaxy. If one of the images suffers negligible extinction,
or if the extinction curve is the same throughout the galaxy, then the
relative intensity ratios of any two images as a function of
wavelength is a direct measure of the extinction curve.

As a part  of  the CLASS radio   survey, Myers et  al.  (\cite{Myers})
reported the discovery  of  a new gravitational lens  candidate, CLASS
B1152+199, with an image separation of $\theta =1\farcs 56$ and a flux
ratio of  $3.03\pm 0.03$ at 8.46   GHz.
Spectra obtained with the Keck II telescope  revealed a background quasar
at  $z=1.019$ and a  foreground   galaxy at $z=0.439$.  Optical
follow-up   observations 
%performed (under non-photometric  conditions)
%with the COSMIC camera on the  Palomar 5-m telescope 
revealed a bright
unresolved object ($g\simeq 16.5$ and  $i\simeq 16.6$) located at the radio position,
presumably the brighter of the two lensed images.  The non-detection of
the weak image in $g$ and $i$ is indicative of a large color difference
between   the two     images.  If  \object{B1152+199}   is   indeed  a
gravitationally lensed system this could be attributed to differential
extinction  caused by dust  in the  lens, making \object{B1152+199}  a
well-suited  system  for studying the extinction law of the lensing
galaxy.

\section{Observations and reduction}
We have performed multiwavelength  $UBVRIz^\prime$ photometry on data
of \object{B1152+199} obtained  with  the  ALFOSC instrument   at  the
2.56-m  Nordic  Optical    Telescope (NOT).  The ALFOSC instrument   is equipped with a Loral/Lesser   CCD
with an  approximate gain  of  $q=1\, e^-/\rm{ADU}$  and $\rm{RON}=6\,
e^-/\rm{pixel}$ and a pixel scale  of $0\farcs188 /\rm{pixel}$.   
The observations   were performed within the span of a few days to  avoid
time delay and   variability effects.  
The observing log is given  in Table~\ref{tabel1}.  The conditions
were photometric with average seeing ranging
from $0\farcs 8$ in  the  $z^\prime$ band to $1\farcs   3$ in the  $U$
band.   The CCD frames  were reduced (overscan subtracted, trimmed and
flatfielded)    using  an IRAF-based   pipeline.\footnote{developed by
Andreas O. Jaunsen}
 
On June 25 1999 UT, we also obtained photometric $UBVRI$ observations of 
standard stars in M92. Using these and calibrated
photometry by Davis (private communication) we fitted the
transformation equations to 19 stars in M92 to derive the color,
extinction and zero-point terms for the calibration to the
Johnson--Kron--Cousins system.

%%%%%%%%%%%%%%%%%%%%%%%%%%%%%%%%%%%%%%%%%%%%%%%%%%%%%%%%%%%%%%%%%%%
\begin{table}

\begin{tabular}{lllllll} \hline
Date         &     $U$&    $B$&    $V$&    $R$&    $I$&   $z^\prime$ \\ 
UT           &    sec&  sec&  sec&  sec&  sec&  sec \\ \hline
June 23 1999 &       &     &     &  600&  600&      \\
June 25 1999 &   2700& 1100&  400&     &     &  900 \\
July 02 1999 &       &  400&  600&     &     &      \\ \hline
\end{tabular}
\caption[]{{Observing log giving the dates of observation and the total 
exposure times for each night}}
\label{tabel1}

\end{table}
%%%%%%%%%%%%%%%%%%%%%%%%%%%%%%%%%%%%%%%%%%%%%%%%%%%%%%%%%%%%%%%%%%% 

\section{Deconvolution of \object{B1152+199}}

The two QSO images together with the lens galaxy were deconvolved with the MCS deconvolution algorithm
(Magain et al.~\cite{Magain}).  This method  is based on the principle
that the resolution of a deconvolved image must be compatible with its
sampling, which is limited  by the Nyquist frequency. 
In order to improve the sampling of the images we adopted a pixel size
in the deconvolved  images as  half the  pixel  size  of the  original
frames ($0\farcs188/2 = 0\farcs094$). (See Burud et
al.~\cite{burud} for more details on the implementation of the MCS algorithm).  
%The  deconvolved image is  decomposed  into a number of   point
%sources for  which the program  returns the  positions and
%intensities,  plus a  deconvolved numerical  background. 

%The MCS algorithm can  be  used to deconvolve a single stacked  image
%or    to simultaneously   deconvolve   several individual images  of a field. 

%The advantage   of the latter   is to derive the   optimally
%constrained  deconvolved  frame   which   is simultaneously
%compatible with  several  different images of a   given object.  
%This  results  in a  more accurate  decomposition of the data than  
%the  deconvolution  of  one  single  combined  frame.  Moreover,
%applying the algorithm to many  dithered frames leads to a deconvolved
%image with an improved sampling.

\begin{figure}
  \resizebox{\hsize}{!}{\includegraphics{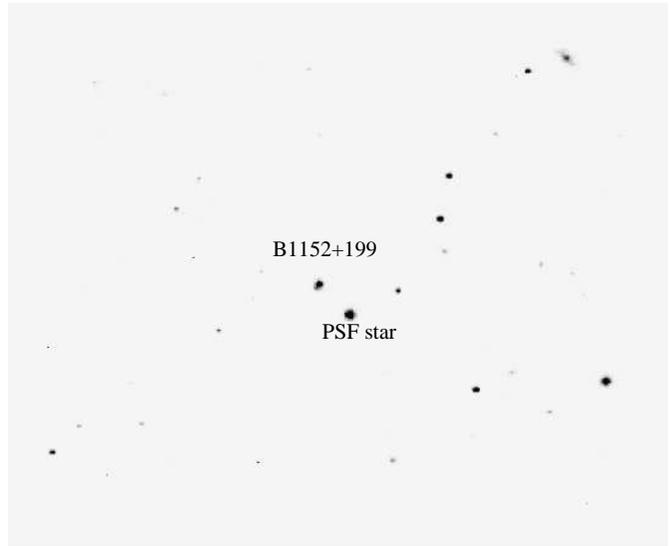}}
  \caption{$160 \arcsec \times 140 \arcsec$ finding chart for
    \object{B1152+119}. North is up, East is to the left. The
    star used to model the point-spread function is indicated}
  \label{finding}
\end{figure}

%Simultaneous   deconvolution  was applied   to the  $B$,  $V$,  $R$,
%$I$  and $z^\prime$ band frames, whereas for the $U$ band, only the
%stacked frame was deconvolved since  the  signal was  very weak in
%each  individual frame.  
In   order     to successfully deconvolve   the   images,  the
point-spread  function (PSF) must    be bright  and located   close to
\object{B1152+199} to minimize the photon noise and
to avoid PSF variation as a function of the  position on the chip.  We
used  the bright star 13$\arcsec$  South West of \object{B1152+199} as
a PSF star (Fig.~\ref{finding}).
%%%%%%%%%%%%%%%%%%%%%%%%%%%%%%%%%%%%%%%%%%%%%%%%%%%%%%%%%%%%%%%%%%%%
%%%%%%%%%%%%%%%%%%%%%%%%%%%%%%%%%%%%%%%%%%%%%%%%%%%%%%%%%%%%%%%%%%%
Careful examination revealed that this star had a faint red companion. 
By deconvolving this star with another PSF in the field we were able
to resolve the red companion star and subtract it, before using the
bright star as a model for the PSF. We compared the star to other
stars in the field to check that the companion star was correctly subtracted.
The best fitting  positions of the weak QSO image relative
to the strong QSO image returned from the MCS algorithm is given in
Table~\ref{flux}. These yield a smaller separation that expected from
the radio positions (Myers et al.~\cite{Myers}). 
The deconvolution code is very sensitive to PSF variations, and small
residuals from the subtraction of the red companion could  explain the
discrepancies between the fitted positions and the radio position of the weak QSO image. 

%This star was finally compared to two more distant stars in order to check that the companion star was correctly subtracted.
%However, small errors might still be present and since the deconvolution
%code is very sensitive to PSF variations this is probably the reason
%for the large uncertainty in the determination of the position of the weak component (given in  Table~\ref{flux}).

In order to improve the photometry we fixed the position of the weak
QSO image relative to the bright QSO image to the
radio coordinates.
The quality of  the results  was checked   from   the residual
maps, as   explained  by   Courbin et al. (\cite{Courbin}).
During the deconvolution the lens galaxy can be modeled
numerically or by an analytical function (e.g., de Vaucouleurs,
exponential-disk law). The lens galaxy in \object{B1152+199} was well fitted by the
numerical model and an exponential disk law, whereas a de Vaucouleurs
model gave significantly bad residuals.
The  best fitting exponential  disk law was centered at
$(\Delta   \alpha  ,\Delta     \delta)= (0\farcs29\pm    0\farcs02
,-0\farcs89 \pm 0\farcs04)$ relative to the bright QSO image. 

\begin{center}
\begin{figure}
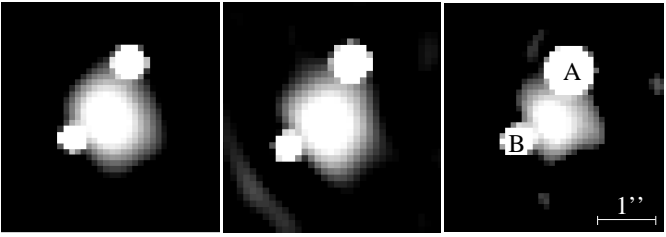

  \resizebox{\hsize}{!}{\includegraphics{H2006.f2}
  \hspace{1mm}
  \includegraphics{H2006.f3}
  \hspace{1mm}
  \includegraphics{H2006.f4}}
  \caption{Deconvolved images of \object{B1152+199} in the (from left
  to right)  $z^\prime$, $R$ and $I$ bands.
The resolution (FWHM) is 0\farcs28 in the $R$ band image and 0\farcs19 in the
$I$ and $z^\prime$ band images.  The bright QSO component is labeled ``A'' and 
the faint component is labeled ``B''  
}
  \label{Deconv}
\end{figure}
\end{center}

Both components of \object{B1152+199} were resolved in the 
$V$, $R$, $I$ and $z^\prime$ bands (Fig.~\ref{Deconv}).  
Following Myers et al. (\cite{Myers}) we will refer to the bright component as
``A'' and  the weak component as ``B''.  
In the $U$ and $B$ bands  we failed to detect  the B image due to the
increasing size  of  the seeing disc with  shorter  wavelength and the
large    differential   extinction  of     the   system (see
Sect.~\ref{extinc}).  
The  lens   galaxy  was  detected  in  the  $R$,  $I$  and
$z^\prime$ bands.  We performed aperture photometry of the galaxy on
the deconvolved images with an aperture diameter of  $2\farcs26$.
 The $U$, $B$, $V$, $R$ and $I$ band magnitudes of the A and B images  and the galaxy (G) were
calibrated (Table~\ref{standard}) whereas  no standard  photometry was
available in the $z^\prime$ band.

%The  galaxy was best  fitted  by an exponential  disk law  centered at
%$(\Delta   \alpha  ,\Delta     \delta)= (0\farcs29\pm    0\farcs02
%,-0.89\arcsec \pm 0.04\arcsec)$ relative to the A image. 
Additional extended signal was observed just South East of  the B image (notably
in the $I$ band). We discuss this further in Sect. \ref{sec:psf}.

\section{PSF photometry of \object{B1152+199}}
\label{sec:psf}
In addition to the deconvolution we 
performed PSF photometry on the system with the IRAF DAOPHOT package
(Stetson \cite{stetson}). As PSF star we used the same star as in the
deconvolution, with the faint red companion removed with the IRAF task
substar. Again we fixed the position of the B image relative to
the A image on the the radio position to ensure the correct
separation.
The results from this analysis is not as robust as the results from
the  deconvolution since it does not simultaneously fit a model for the
galaxy, but the differences between the results of the
two approaches is indicative of the magnitude of the systematic photometric uncertainties. 

\begin{center}
\begin{figure}
  \resizebox{\hsize}{!}{\includegraphics{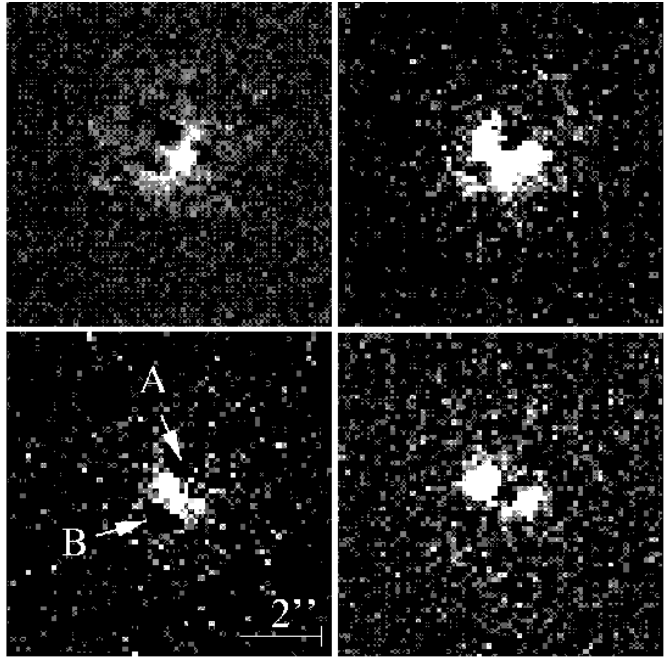}}
  \caption{{PSF-subtracted $z^\prime$ (top left), $I$ (top right), $R$
      (bottom left), and $B$ (bottom right) images}}
  \label{Clean}
\end{figure}
\end{center}

The two QSO images were resolved in the $B$, $V$, $R$, $I$ and
$z^\prime$ bands. The results from the PSF photometry is given in Table
\ref{flux}. 
After subtracting the scaled PSFs we detected the
lensing galaxy  in the  $B$, $R$, $I$ and $z^\prime$ band
residual images (Fig. \ref{Clean}). 
The $I$ and $z^\prime$ band images indicates that the extended flux South East of the
B image, detected in the deconvolved images,  stems from
the lensing galaxy.  
The light from  the B component passes through the galaxy close to
its center, and in the deconvolution process some of the light from the galaxy may have been attributed to the B image.
  
To test the results from the deconvolution we performed aperture
photometry with the SExtracter software (Bertin \&
Arnouts~\cite{bertin}) on the images cleaned with DAOPHOT.
The residual images in Fig. \ref{Clean} show artifacts of the PSF
subtraction (seen as circular dark spots). These affect the aperture photometry
which should therefore be considered rather uncertain. 
In the $I$ and $B$ bands we obtain $I=19.43\pm0.03$ and $B=20.24\pm0.02$
with an aperture diameter of $2\farcs26$. 
In the $R$ band the signal is rather weak, and thus in the outer parts
dominated by the PSF subtraction artifacts. We can however obtain a color
by comparing the $R$ and $I$ flux within an aperture of $1\farcs13$.
This yields  $R-I=0.80\pm0.05$. 
The color of the galaxy from the deconvolution  
$(R-I)_{\rm{MCS}}=0.43 \pm 0.06$ or from PSF photometry
$(R-I)_{\rm{PSF}}=0.80\pm 0.05$ are consistent with the lensing galaxy
being a irregular or late type galaxy at $z=0.439$
(Fukugita et al.~\cite{fukugita}).
By comparing the photometric results from the two approaches we
estimate the  magnitude of the systematic photometric errors due to
the limited quality of the data to be in the order of 0.4 mag for
galaxy magnitudes and 0.2 mag for the QSO magnitudes.

%%%%%%%%%%%%%%%%%%%%%%%%%%%%%%%%%%%%%%%%%%%%%%%%%%%%%%%%%%%%%%%%%%%

%%%%%%%%%%%%%%%%%%%%%%%%%%%%%%%%%%%%%%%%%%%%%%%%%%%%%%%%%%%%%%%%%%%

%%%%%%%%%%%%%%%%%%%%%%%%%%%%%%%%%%%%%%%%%%%%%%%%%%%%%%%%%%%%%%%%%%%
\begin{table}

\begin{tabular}{lllll} \hline
Filter &  (B$-$A)$_{\rm{MCS}}$& $\Delta\alpha$ &$\Delta \delta$ &(B$-$A)$_{\rm{PSF}}$ \\ 
         &  mag  & $\arcsec$ & $\arcsec$ &mag \\ \hline 
$B$      &       &           &              &5.84 $\pm$ 0.51\\
$V$      &  4.42 $\pm$ 0.02&  0.88 &$-1.19$&4.44 $\pm$ 0.12  \\
$R$      &  3.87 $\pm$ 0.06&  0.78 &$-1.29$&3.67 $\pm$ 0.06  \\
$I$      &  2.98 $\pm$ 0.01&  0.86 &$-1.22$&2.69 $\pm$ 0.03 \\
$z^\prime$ &  2.61 $\pm$ 0.04&0.94 &$-1.20$&2.29 $\pm$ 0.04 \\ \hline  
\end{tabular}

\caption[]{{Columns  2--4 give the flux ratios and astrometry for
       \object{B1152+199} from the deconvolution. 
$\Delta\alpha$ and $\Delta
   \delta$ are the fitted positions  of the
    B image relative to the A image. The rms uncertainties in
    the positions are approximately $0\farcs05$ in $\Delta\alpha$ and
    $0\farcs01$ in $\Delta\delta$. The positions agree reasonably well with
    the radio positions $(\Delta\alpha, \Delta\delta)= (0\farcs935 \pm0\farcs005
, -1\farcs248\pm0\farcs005$), although the separation is slightly
    smaller. This is a well-known effect due to the entanglement of the B
    image and galaxy flux.
Column 5 gives the flux ratios from the PSF photometry }}
\label{flux}
\end{table}
%%%%%%%%%%%%%%%%%%%%%%%%%%%%%%%%%%%%%%%%%%%%%%%%%%%%%%%%%%%%%%%%%%%

%%%%%%%%%%%%%%%%%%%%%%%%%%%%%%%%%%%%%%%%%%%%%%%%%%%%%%%%%%%%%%%%%%
\begin{table}
\begin{tabular}{lllll} \hline
     &FWHM   &A             &B      &G             \\ \hline
$U$ &$1\farcs22$     &16.75$\pm0.04$&       &                     \\
$B$ &$1\farcs09$     &17.00$\pm0.02$&       &                     \\
$V$ &$1\farcs02$     &17.28$\pm0.03$&$21.73\pm0.03$&              \\
$R$ &$1\farcs00$     &16.42$\pm0.02$&$20.20\pm0.04$&$19.61\pm0.04$\\
$I$ &$0\farcs94$     &16.38$\pm0.02$&$19.26\pm0.03$&$19.18\pm0.04$\\
$z^\prime$   &$0\farcs83$           &     &       &         \\ \hline
\end{tabular}

\caption[]{{Photometry (mag) of image A, B and the lensing galaxy. The
    uncertainties are $1\sigma$ standard deviations, including
    standard-star zero point, calibration and photometry uncertainties }}
\label{standard}
\end{table}

%%%%%%%%%%%%%%%%%%%%%%%%%%%%%%%%%%%%%%%%%%%%%%%%%%%%%%%%%%%%%%%%%%%

\section{Galaxy extinction}
\label{extinc}

Significant differential  extinction of the B image  relative to the A
image was observed (Tables~\ref{flux} and \ref{standard}).  The A image,
however, does  not  appear  to  be significantly   reddened.  In  what
follows we assume that the A image is unaffected by extinction by dust
in the  lensing galaxy. 
%These   conditions make \object{B1152+199}  an ideal system for measuring the extinction curve of the lensing galaxy.
The Galactic extinction curve is fairly well known, with $A_{\lambda}$
well parameterized  as a   function of  a single  parameter,  $R_{V}$,
defined  as the ratio between  the $V$  extinction,  $A_V$, and the color
index    $E(B-V)$     (Cardelli  et    al.     \cite{Cardelli};
Fitzpatrick \cite{fitz}).  This
parameterization is  valid over  the  observed $R_V$ values ranging
from $2.5$  to $6$,   with a  mean     (canonical) Galactic  value  of
$R_V=3.1$. Large  $R_V$ give flat  UV  extinction curves due  to large
grains whereas   small $R_V$  and   small grains  lead  to  large   UV
extinctions.

The variation of the flux ratio with wavelength  directly measures the
extinction curve of the galaxy. We have assumed that the intrinsic
flux ratio is given by the radio flux ratio (B$-$A=1.21). 

%%%%%%%%%%%%%%%%%%%%%%%%%%%%%%%%%%%%%%%%%%%%%%%%%%%%%%%%%%%%%%%%%%%%%%%%%%%
The extinction  curve of \object{B1152+199} was well fitted  by a
Galactic extinction law, using either the flux ratios from the
deconvolution or from the PSF photometry. However the best fitting values
of the parameters $R_V$ and $E(B-V)$ differ somewhat. We first discuss the fit 
 to the deconvolution data.
In the deconvolved  $U$ and $B$ band images the galaxy and the B
image were too weak to be separated. We can however obtain upper
limits on the B image flux by using the total flux of the galaxy and the B image. These are $B_{\rm{galaxy+B}}=20.37$ and
$U_{\rm{galaxy+B}}=20.45$. 
The best fitting restframe values of the  parameters to the deconvolution flux ratios are 
 $R_V=2.04 \pm 0.04$ and $E(B-V)=0.89\pm 0.02$ (Fig.~\ref{ext},
 solid curve). The predicted magnitude of the B image from the
 observed A image magnitude  and  $R_V=2.04$ are $B_{\rm{B}}=21.69$
and $U_{\rm{B}}=23.27$. Furthermore the inferred $B_{\rm{galaxy+B}}$
agree well with the $B$ band magnitude of the galaxy inferred from the PSF
photometry (Sect. \ref{sec:psf}).
It is therefore likely that the  upper
limits presented above are  dominated by the  galaxy flux rather  than
flux from the B image.  

The PSF photometry detected the B component in the $B$ band which gives
 us another point on the extinction curve. The best fitting restframe
 values of the  parameters to the PSF photometry flux ratios are
 $R_V=1.32 \pm 0.02$ and $E(B-V)=1.13\pm 0.01$ (Fig.~\ref{ext},
 dashed curve).
The extinction curves inferred from the two approaches are
comparable. As discussed in Sect. \ref{sec:psf} the systematic uncertainties in the photometry is of the order 0.2 magnitudes. This
uncertainty is illustrated by the different slopes of the
extinction curves derived from the two approaches.
We conclude that the extinction curve of the lensing galaxy of \object{B1152+199} is
fitted by the Galactic extinction law and put the following constrains
on the parameters: $\protect{1.3 \lesssim R_V \lesssim 2.1}$ and $0.9
 \lesssim E(B-V) \lesssim 1.1$.

%%%%%%%%%%%%%%%%%%%%%%%%%%%%%%%%%%%%%%%%%%%%%%%%%%%%%%%%%%%%%%%%%%%%%%%%%

%The extinction  curve of \object{B1152+199} was well fitted  by a
%Galactic extinction law, with restframe values of $R_V=2.04 \pm 0.04$
%and $E(B-V)=0.89\pm 0.02$ (Figure~\ref{ext}).  The predicted magnitude
%of   the  B image    from the observed    A image   magnitude  and 
%$R_V=2.04$ are $B_{\rm{B}}=21.69$
%and $U_{\rm{B}}=23.27$. It   is therefore likely that the  upper
%limits presented above are  dominated by the  galaxy flux rather  than
%flux from the B image.  
%As noted in Section~\ref{results} the B image flux might be
%contaminated by flux from the lensing galaxy.  Since the galaxy is
%blue the effect of the contamination is largest in the blue wave
%bands. Correction for this effect would steepen the extinction curve,
%i.e, lower the value of $R_V$. The inferred value of $R_V$ thus
%represents an upper bound to the real value. 

If the reddening is due  to dust extinction in
the lens galaxy we can derive the column  density of neutral hydrogen from the
observed $E(B-V)$ by assuming a gas to dust ratio for the galaxy.  
For the Milky \protect{ Way
$\left < \rm{N}(\ion{H}{I})/E(B-V)  \right
  >_{\rm{MW}}=4.93 \times 10^{21}$\,cm\,$^{-2}$~mag$^{-1}$,}
(Diplas   \&  Savage \cite{Diplas})  
while the value  for Damped Lyman$\alpha$  Absorbers  (DLAs) may be up to  20
times  larger (Fall \& Pei \cite{Fall}).   With  the inferred $E(B-V)
\approx 1$
for the   lens   galaxy: \protect{$\rm{N}(\ion{H}{I})_{\rm{{\object{B1152+199}}}}=k
  \times 10^{21}$\,cm\,$^{-2}$,  where  $5 \lesssim k
\lesssim 100$  is  the
parameterization of   the range in  the   observed $\left  <
  \rm{N}(\ion{H}{I})/E(B-V) \right >$ in different systems. Thus, the
lensing galaxy of  \object{B1152+199}
would qualify as a DLA (N(\ion{H}{I})$>2\times 10^{20}$\,cm\,$^{-2}$) regardless of the value of $k$. 
%%%%%%%%%%%%%%%%%%%%%%%%%%%%%%%%%%%%%%%%%%%%%%%%%%%%%%%%%%%%%%%%%%%%
\begin{figure}
  \resizebox{\hsize}{!}{\includegraphics{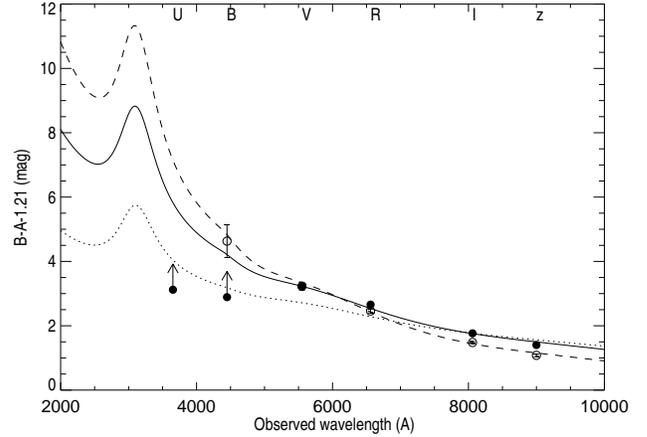}}
  \caption{Extinction curve of the lensing galaxy of \object{B1152+199}, fitted by a Galactic
  extinction law. The best fitting parameters to  the deconvolution data (filled symbols) are  
    $R_V=2.04 \pm 0.04$ and $E(B-V)=0.89 \pm 0.02$ (solid curve). The $U$
    and $B$ band points are lower limits (see text). The best fitting
    parameters to the PSF photometry data (open symbols) are $R_V=1.32
    \pm 0.02$ and $E(B-V)=1.13 \pm 0.01$ (dashed curve). For comparison we
    have overplotted the Galactic ($R_V=3.1$) extinction curve (dotted
    curve) }
  \label{ext}
\end{figure}
%%%%%%%%%%%%%%%%%%%%%%%%%%%%%%%%%%%%%%%%%%%%%%%%%%%%%%%%%%%%%%%%%%%

%%%%%%%%%%%%%%%%%%%%%%%%%%%%%%%%%%%%%%%%%%%%%%%%%%%%%%%%%%%%%%%%%%%
\begin{table}

\begin{tabular}{lllllllll} \hline
$\Delta\alpha$   &$\Delta\delta$  &e     &P.A.&$\rm{r_{core}}$ &$\rm{r{_{cut}}}$ &$\sigma$&delay\\ 
\arcsec      & \arcsec    &1$-$b/a &deg & kpc       & kpc      & km/s   &days \\ \hline
0.56        &$-0.87$       &0.21  &21  &0.02       & 5.42     & 253    &59 \\ \hline
 
\end{tabular}
\label{model}
\caption[]{{Simple lens model parameters for \object{B1152+199} (cf.~Fig.~\ref{modfig}). $\Delta\alpha$, $\Delta\delta$ are the
    positions of the lensing galaxy relative to the A component. 
    The position of the galaxy and the QSO images along with their flux
    ratios were held constant in the fitting procedure, while the time delay
    was calculated from the best fit to the remaining parameters.
    $\Omega =0.2$, $\Lambda =0$, $H_0=65$ km s$^{-1}$ Mpc$^{-1}$ were assumed}}
\end{table}

%%%%%%%%%%%%%%%%%%%%%%%%%%%%%%%%%%%%%%%%%%%%%%%%%%%%%%%%%%%%%%%%%%% 

\section{A simple model of \object{B1152+199}}
 
A simple model with a dark  halo, represented by  
\begin{eqnarray}
\nonumber
\Sigma(R)={\Sigma}_0\frac{r_{\rm{core}}r_{\rm{cut}}}{r_{\rm{cut}}-r_{\rm{core}}}\left
( \frac{1}{(r_{\rm{core}}^2+R^2)}-\frac{1}{(r_{cut}^2+R^2)} \right ),
\end{eqnarray}
as the  dominating mass was fitted to the data (Kneib et al. \cite{kneib}).
Although the observational  constraints available from our  data were
too  poor to  perform detailed   modeling, a  set of best-fitting
parameters could be derived   for the system  (see Table~\ref{model}).
We used the radio  positions and  flux  ratios for the  two QSO components
while we experimented with different   positions for the lens  galaxy.
We found that the two-image configuration could not be reproduced with
the  galaxy in  the position  found   from the  deconvolved images. 
An improved fit was obtained by placing  the galaxy  closer  to the  B
component. The position was chosen to be consistent with what is found
in the PSF-subtracted   images. 
%In  simple models  the   weaker of two components is expected  to be
%closer to the   center of mass  than the stronger one. 
This position means that the B component passes 
through the  galaxy while the A  component  passes outside the galaxy,
which   is   consistent  with    the  large   observed    differential
extinction. The model predicts a time delay of 59 days.
  In Fig.~\ref{modfig} we  show the  critical and caustic
curves of the lens model overlaid the deconvolved $R$ image.

%%%%%%%%%%%%%%%%%%%%%%%%%%%%%%%%%%%%%%%%%%%%%%%%%%%%%%%%%%%%%%%%%%%

\begin{center}
\begin{figure}
  \resizebox{\hsize}{!}{\includegraphics{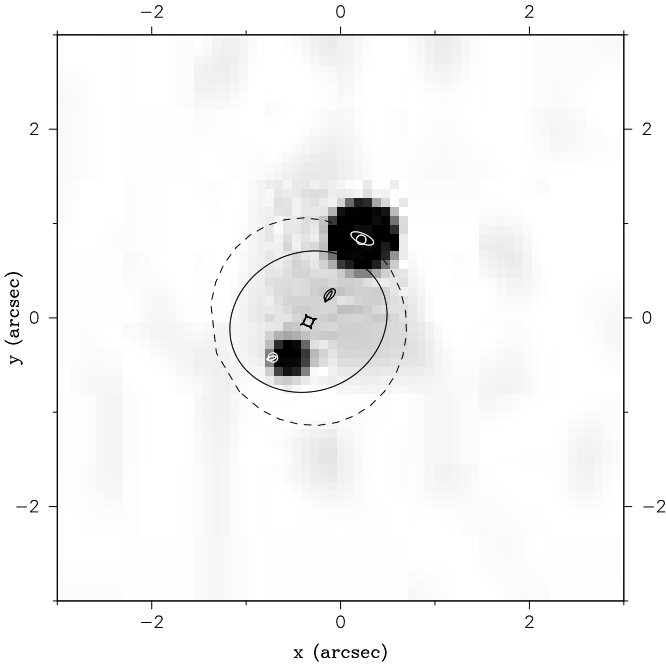}}
  \caption{Critical (solid) and caustic (dashed) curves from the model, overlaid the deconvolved $R$ image} 
  \label{modfig}
\end{figure}
\end{center}

%%%%%%%%%%%%%%%%%%%%%%%%%%%%%%%%%%%%%%%%%%%%%%%%%%%%%%%%%%%%%%%%%%%

\section{Discussion}
Extinction curves have been derived for the Galaxy and the Magellanic Clouds, 
mainly through multi-color photometry of individual stars,
for which the intrinsic luminosities is known from their spectral types
(Cardelli et al. \cite{Cardelli}).  The extinction curves of the LMC
and SMC can be fitted  by the Galactic extinction law 
(see Mathis~\cite{mathis}; Fitzpatrick~\cite{fitz}). The main 
difference is the 2175~\AA\ bump (cf. Fig.~\ref{ext}) which is not observed in the LMC and SMC. 
At distances $\gtrsim$ 10 Mpc study of individual stars is difficult and
other methods must be considered.
In this paper we have shown that accurate extinction  curves of high-$z$ galaxies
can be determined by studying gravitationally lensed systems. 
We have found that the extinction curve of \object{B1152+199} is well
fitted by the Galactic extinction law. It is interesting to note that
the inferred  $1.3 \lesssim R_V \lesssim 2.1$ in
principle is outside the validity of the parameterization of Galactic
extinction curve by Cardelli et al.~(\cite{Cardelli}). 
Note also that the differential extinction of
\object{B1152+199} $0.9 \lesssim E(B-V) \lesssim 1.1$, is one of the largest observed to
date (Falco et al.~\cite{Falco}).   

%The quality of our data was not good enough to sample the extinction curve
%in the $B$ and $U$ band. This is unfortunate since these correspond to
%restframe UV radiation which is highly sensitive to dust extinction,
%and thus strongly constrains the shape of the extinction curve.  
%A detection of the B image in the $U$ band would also
%have allowed a possible detection of the 2175 \AA\ 
%bump.  The evidence for a Galactic extinction curve is intriguing, but
%$U$ and $B$ band observations with longer exposure times and better seeing
%are needed to rule out other extinction curves. 

The evidence for a Galactic extinction curve is intriguing, but
$U$ and $B$ band observations with longer exposure times and better seeing
are needed to rule out other extinction curves. These will sample
the restframe UV radiation which is highly sensitive to dust extinction,
and thus strongly constrain the shape of the extinction curve, and
allow for the  possible detection of the 2175 \AA\ bump.

The predicted time delay of 59 days makes \object{B1152+199} an
interesting candidate for monitoring.

%Using this method, a survey of extragalactic extinction curves could be
%made by observing gravitationally lensed systems with
%differential extinction at a range of redshifts and morphologies. The
%results of such an survey would provide important clues to galaxy evolution
%and dust evolution. Simultaneous multiwaveband photometric and
%spectroscopic observations should be made for each candidate. This
%would allow for the correction for the main sources of systematic
%error, namely variational and microlensing effects.

%A survey of extragalactic  extinction curves for lenses  at a range of
%redshifts and morphologies would  provide  important clues  to  galaxy
%evolution and  dust evolution.  Simultaneous multiwaveband photometric
%and spectroscopic observations should be made for each system to allow
%correction for the main sources of systematic error, namely variations
%due to time delay and microlensing effects.

\begin{acknowledgements}
We thank Andreas O. Jaunsen for providing the pipeline for the reduction,
Jean-Paul Kneib for providing his Lens Tool software for modeling the
system, and Lars Freyhammer for performing the observations.
We thank Kirsten K. Knudsen and Anja C. Andersen for valuable 
discussions. IB is supported in   part  by  contract
ARC94/99-178 ``Action   de Recherche Concert\'ee  de  la  Communaut\'e
Fran\c{c}aise (Belgium)'' and P\^ole d'Attraction  Interuniversitaire,
P4/05 \protect{(SSTC, Belgium)}. 
The data presented here have been taken using ALFOSC, which is owned
by the Instituto de Astrofisica de Andalucia (IAA) and operated at the
Nordic Optical Telescope under agreement between IAA and the Astronomical Observatory of Copenhagen.
This work was supported by the Danish Natural Science Research Council
(SNF).
\end{acknowledgements}

\end{document}